\documentclass[10pt,prb,nobibnotes,showpacs,showkeys,superscriptaddress,maxnames=6]{revtex4-1}
\usepackage{geometry}
\usepackage{amsfonts,amsmath,amssymb,graphicx, placeins, bm}
\usepackage{setspace,color}

\usepackage{ae,aeguill}
\usepackage[utf8]{inputenc}    
\usepackage[T1]{fontenc}
\newcommand{\cor}[1]{#1}

\newcommand{\ddi}{\hat{\mu}_i\hat{\mu}_j -3 (\hat{\mu}_i \hat{r}_{ij})(\hat{\mu}_j \hat{r}_{ij})}

\newcommand{\ga}{\gamma}  \newcommand{\be}{\beta}    \newcommand{\de}{\delta} 
\newcommand{\la}{\lambda} \newcommand{\al}{\alpha}   \newcommand{\Ga}{\Gamma}
\newcommand{\gD}{\Delta}  \newcommand{\ep}{\epsilon} \newcommand{\s}{\sigma}
\newcommand{\om}{\omega}     
 \newcommand{\tend}{\rightarrow}
\newcommand{\equa}[1]{\begin{eqnarray} \label{#1}}
\newcommand{\auqe}{\end{eqnarray}}
\newcommand{\equab}[1]{\begin{widetext}\begin{eqnarray} \label{#1}}
\newcommand{\auqeb}{\end{eqnarray}\end{widetext}}
\newcommand{\tab}[1]{\begin{tabular}{#1}}
\newcommand{\bat}{\end{tabular} \\ }
\newcommand{\blanc}{\hskip 0.10\textwidth}

\begin {document}
  \title {Quantifying time in Monte Carlo simulations: application to relaxation processes and AC susceptibilities 
           of magnetic nanoparticles assemblies.}
  \author {A. Morjane}
  \affiliation {PROMES-CNRS (UPR-8521), Université de Perpignan Via Domitia,
  Rambla de la thermodynamique, Tecnosud, 66100 Perpignan, France. }
  \affiliation{ICMPE, UMR 7182 CNRS and UPE 2-8 rue Henri Dunant 94320 Thiais, France.}
  \author {J.-G. Malherbe}
  \email [e-mail address: ] {malherbe@u-pec.fr}
  \affiliation{ICMPE, UMR 7182 CNRS and UPE 2-8 rue Henri Dunant 94320 Thiais, France.}
  \author{Juan J. Alonso}
  \email [e-mail address: ] {jjalonso@uma.es}
  \affiliation{F\'{\i}sica Aplicada I, Universidad de M\'alaga, 29071 M\'alaga, Spain}
  \affiliation{Instituto Carlos I de F\'{\i}sica Te\'orica y Computacional,  Universidad de M\'alaga, 29071 
               M\'alaga, Spain}
  \author {F. Vernay}
  \affiliation {PROMES-CNRS (UPR-8521), Université de Perpignan Via Domitia,
  Rambla de la thermodynamique, Tecnosud, 66100 Perpignan, France. }
  \author {V. Russier}
  \email [e-mail address: ] {vincent.russier@cnrs.fr}
  \affiliation{ICMPE, UMR 7182 CNRS and UPE 2-8 rue Henri Dunant 94320 Thiais, France.}
  \begin {abstract}
    \vskip 0.05\textheight

The study of the response of  magnetic nanoparticles (MNP) assemblies to an
external alternating magnetic field is of great interest for applications such as 
hyperthermia. The key quantity here is the complex susceptibility and its
behavior in terms of temperature and frequency. From a theoretical point of view
it can be obtained by Monte Carlo (MC) simulation with the time quantified Monte 
Carlo (TQMC) method if a physical time is associated with the MC step. Here we revisit this
method by showing that the time unit can be derived from the MC stochastic
process of the isolated particle. We first 
\cor {obtain a MC unit of time from} 
the relaxation of the
system at fixed temperature. Then this 
\cor {unit of time} 
is used to compute complex susceptibilities. We show that it is now possible to match 
the TQMC results with  
actual experimental results regarding  frequency dependent in
phase susceptibilities
\cor {and quantify the unit of time in seconds.} 
Finally we show that the time unit obtained 
for the isolated particle remains valid when considering interacting
particles such as the Heisenberg coupling or dipole dipole
interactions.

  \end {abstract}
  \maketitle
%

\section {Introduction}
\label   {intro}
Magnetic nanoparticles (MNP) assemblies still arouse a great interest on the experimental and 
theoretical points of view due to the quite large field of applications~\cite{bedanta_2009}. 
In particular the dynamic properties, either stemming from the blocking or Néel process or resulting 
from the application of an oscillating external field, are of great importance such as the 
induced heat transfer to the embedding medium giving rise to the magnetic hyperthermia 
with applications in cancer therapy~\cite{abenojar_2016,ovejero_2021,gavilan_2023}
or catalysis~\cite{bordet_2016} for instance.
In the framework of mesoscopic scale models such as the family of effective one spin models,
where the MNP are represented as homogeneously polarized single domain objects, the
simulation of dynamic properties of magnetic nanoparticles assemblies can be performed through
different methods or approaches. Many of them start from the stochastic LLG equation namely 
the deterministic Gilbert equation to which the thermal fluctuations are included through a 
fluctuating field, leading to the Fokker-Planck equation for the spin orientation probability 
distribution.
On an other hand, the Monte Carlo simulation method (MC), besides the two-states 
scheme~\cite{andersson_1997,brinis_2014} and the kinetic MC~\cite{tan_2014}, has been adapted in order 
to quantify the Monte Carlo step to an actual time step 
\cor {measured in seconds}.  
After the dynamical Monte Carlo scheme used in~\cite{garcia-otero_2000} where dynamics are introduced {\it via}
a temperature independent constraint on the move amplitude of the moments, the actual
pioneering works in this direction are the  so-called time quantified Monte Carlo (TQMC) 
method by Nowak {\it et al.}~\cite{nowak_2000}, and extended by Cheng {\it et al.}~\cite{cheng_2006}.
As a general rule, the different applications of the TQMC concern  ensembles of MNP fixed in position 
and either non interacting~\cite{nowak_2000,chubykalo_2003,billoni_2007,melenev_2012}, namely isolated MNP, 
or including short range Heisenberg~\cite{cheng_2006} 
or dipolar interactions (DDI)~\cite{stariolo_2008,serantes_2010,fu_2018}.
The starting point is the mapping of the time step from the stochastic LLG
equation to the one of 
the Monte Carlo step through a temperature dependent constraint on the amplitude of the moments 
displacements in the MC sampling  assuming the heat-bath sampling~\cite{nowak_2000,cheng_2006,chubykalo_2003}. 
Melenev {\it et al.}~\cite{melenev_2012} used another method for the mapping of the TQMC scheme by 
introducing the number of MC steps corresponding to the relaxation time of the Néel process as a MC time scale,
and focused on the superparamagnetic relaxation of non interacting NP.
Billoni {\it et al.}~\cite{billoni_2007} used the TQMC in order to study the behavior of switching time of an isolated
MNP in the Stoner-Wolfhart model. 
Stariolo {\it at al.}\cite{stariolo_2008} used the TQMC to calculate both the relaxation time and the autocorrelation
function of an ensemble of NP on a 2-D square lattice, including DDI. 
More recently, this method was used for instance in~\cite{serantes_2010,fu_2018} for modeling the influence of DDI on 
the hyperthermia properties of MNP assemblies.
Stricto-sensu, the TQMC leads to a mapping of the MC step on an actual time scale but this is related to the 
gyromagnetic factor $\gamma$ and the damping factor $\al$. 
However, the precise knowledge of $\al$ for magnetic nanoparticles is missing (see for instance~\cite{back_1999})
and this makes difficult the correspondence between the TQMC step and the actual time scale. 
This may be compared to the usual correspondence made in the FC/ZFC protocol 
between the relaxation time at the blocking temperature and the measuring time where the prefactor of the Arrhenius
form of the Néel relaxation time is generally estimated with a large
uncertainty  (10$^{-8}$ to 10$^{-10}$\;s).
Hence a convenient determination of the time scale through the TQMC is still missing as was mentioned 
in~\cite{hovorka_2014}. 
For this reason, in Refs.~\cite{serantes_2010,serantes_2012} the true time
scale is not made explicit
but instead stated as directly proportional to the MC step (MCS) once the temperature dependent constraint 
mentioned above is applied but in the framework of the Metropolis sampling. 
A similar recipe for performing dynamical MC from Metropolis sampling was already introduced
in~\cite{dimitrov_1996,l-wang_2001,lepadatu_2021} imposing a temperature independent acceptance rate fixed in between 0.3 and 0.4.
Similarly, in~\cite{nguyen_2009a} the usual Metropolis sampling is used with the constraint of a 
constant acceptance ratio to perform MC simulations of the susceptibility in terms of the frequency $\om$ with 
assumed proportionality between $1/\om$ and MCS.
\\ \indent
In this paper, we address this question by demonstrating how 
\cor {a MC unit of time} 
can be obtained from the details of the simulation of the isolated
particles. We show that this 
\cor{unit of time} 
remains valid when dealing with
interacting particles. 
Moreover we show how to extend the TQMC scheme to the Metropolis
sampling instead of the heat-bath one which was used to map the MC time step to the LLG 
one~\cite{nowak_2000,cheng_2006}.
Therefore we explicit a time unit for both the heath bath and Metropolis samplings.
To this aim, 
we propose an alternative way to use the TQMC.
We start from the mapping deduced in~\cite{nowak_2000,cheng_2006} without specifying the values of the gyromagnetic
factor nor the damping factor $\al$, seen as characteristics of the microscopic features of the system
not taken into account in the mesoscopic scale models. 
Instead we leave a free parameter for the correspondence between
the MC step and the actual time step. 
First we demonstrate  that the TQMC step indeed behaves as a convenient time scale
for both non interacting and interacting assemblies of MNP represented by dipolar hard spheres plus uniaxial 
anisotropy (the so-called effective one spin model). 

Then we show from the Néel relaxation process of non interacting MNP
that we recover for the relaxation time in terms of temperature 
$T$ and anisotropy barrier height in $k_BT$ unit, $\s=K_{u}/k_BT$ either the 
Brown~\cite{brown_1963} or the Bessais {\it et al.}~\cite{bessais_1992} form
by fitting a multiplicative constant. This is interpreted as a 
first step to quantify the TQMC scheme.
Then we propose a mapping leading to a correspondence 
\cor{of the MC time} 
with the true time scale through the behavior of 
the real part of the susceptibility with respect to the frequency of real systems which are standard
quantities considered in experiments.

This paper is organized as follows. After presenting the model in Section~\ref{model}, 
the TQMC is revisited  in Section~\ref{method} by defining 
\cor{a unit of time} 
from the details of the
simulation in the sense that it is sampling choice independent. In Section~\ref{results} this
time scale is tested first for non interacting case. Then we  show  how it is possible to map the
simulation results with experimental ones in order to obtain the 
\cor{true} 
time scale. 
\cor {Finally} 
we demonstrate that this 
\cor{unit of time} 
remains valid 
\cor{for} 
interacting systems. The paper ends with a brief conclusion.

%
 \section {Model}
 \label   {model}
 We model an assembly of $N$ spherical MNP in the single domain regime, free of super exchange interactions and 
characterized by a uniaxial magnetocrystalline anisotropy energy (MAE) and monodisperse in size.
 We place ourselves in the framework of the effective one spin model where each single domain MNP is assumed to be
uniformly magnetized with a temperature independent saturation magnetization $M_s$.
 Hence, we
consider a system of dipolar hard spheres of moment $\vec{\mu}_i=\mu_i\hat{\mu}_i$, with $\mu_i=M_{s}v_i$, 
interacting through the usual dipole dipole interaction (DDI), subjected to a 
one-body anisotropy energy, $K_{i}v_i(\hat{n}_i\ldotp\hat{\mu}_i)^2$ and to an external field 
$\vec{H}=H_0\hat{h}$ where $H_0$ may be time dependent.
$K_i$, $\hat{n}_i$ and $v_i$ are the anisotropy constant, the easy axis and the volume of the particle $i$ respectively.
The MNP ensemble is monodisperse with MNP diameter $d$ and moment $\mu$. The Hamiltonian of the system is given by
\equa{beh_1}
  \be H & = & 
        \frac{1}{2} \be \ep_d \sum_{i\neq j} \frac{\ddi}{(r_{ij}/d)^3}
   - \be K_u v(d) \sum_i (\hat{n}_i\ldotp\hat{\mu}_i)^2 + \frac{1}{2} \be \sum_{i\neq j} v_{sr}(r_{ij})
   - \be\mu_0 H_0 \mu \sum_i \hat{h}\ldotp \hat{\mu}_i
 \\ \nonumber   & \textrm{with} &~ \ep_d = \frac{\mu_0}{4\pi}\frac{\mu^2}{d^3} 
\auqe
where $\hat{r}_{ij}$ is the unit vector carried by the vector joining sites $i$ and $j$,
$\be=1/(k_BT)$ is the inverse temperature.
$v_{sr}(r)$ is the short range contribution, taken as the hard sphere potential, 
$v_{sr}(r>d)=0$ and $v_{sr}(r<d)=\infty$.
We introduce a reference temperature, say $T_0$ ($\be_0=1/k_BT_0$) and 
the reduced temperature is chosen as $T^*=T/T_0$
In the following we restrict ourselves to an assembly of fixed MNP with a fixed distribution of easy axes
and accordingly we can drop the short ranged contribution, $v_{sr}$.
Moreover the MNP are located on the nodes of a perfect lattice of either FCC for of BCC symmetry.
Equation~(\ref{beh_1}) is then rewritten as
 \equa{beh_2a}
  \be H  &=&  \frac{1}{T^*} \left( \frac{\la_d}{2} \sum_{i\neq j} \frac{\ddi}{(r_{ij}/d)^3} 
	      - \la_u \sum_i (\hat{n}_i\ldotp\hat{\mu}_i)^2  - h \sum_i \hat{h}\ldotp \hat{\mu}_i  \right )
  \\ \nonumber
    & \la_u & = \be_0 K_uv(d) ~;~ \la_d = \be_0\ep_d ~;~ h = \be_0\mu_{0} \mu H_{0} ~\equiv~ 2\la_{u}\frac{H_0}{H_K}  
 \auqe
which introduces the DDI, MAE and Zeeman coupling constants, $\la_d$, $\la_u$ and $h$,
where $H_K=2K/\mu_{0}M_s$ is the usual anisotropy field.
\\~\\
 We will also consider the case of an effective one spin model where the particles interact through the usual 
Heisenberg interaction of coupling constant $J$, in unit $1/\be_0$, restricted to first neighbour pairs $<ij>$ 
instead of the DDI above. Then, the hamiltonian is given by 
\equa{heis_1}
  \be H  &=&  \frac{1}{T^*} \left( J \sum_{<ij>}\hat{\mu}_i\ldotp \hat{\mu}_j 
   - \la_u \sum_i (\hat{n}_i\ldotp\hat{\mu}_i)^2 - h \sum_i \hat{h}\ldotp\hat{\mu}_i \right)
\auqe
\\~\\
Here, we consider only the ferromagnetic Heisenberg model, $J<0$, and the ordered phase at low temperature 
is either of ferromagnetic (FM) or spin-glass (SG) nature according to the value of $\la_u$~\cite{itakura_2003} 
when the easy axes are randomly distributed.
The simulation box is a cube with edge length $L$, the total number of dipoles is $N$ and the volume fraction
is $\varphi_v=N\pi{}d^3/(6L^3)$.
We consider periodic boundary conditions by repeating the simulation cubic box identically in the 3 dimensions.
The long range DDI interaction is treated through the Ewald summation technique~\cite{allen_1987,weis_1993}. 
%
\section {Method}
\label {method}
 We start from the TQMC scheme as introduced first in Ref.~\cite{nowak_2000} and then improved in Ref.~\cite{cheng_2006},
based on the mapping of the MC time 
to the actual time $t$ of the LLG equation. 
We refer the reader to these works 
for details on how  
this mapping is deduced. 
We just need to recall that it is based on the precise tuning of the aperture of the cone, $R$,
in which the new moment at site $i$ say $\hat{\mu}_i^{(n)}$ is chosen from 
its preceding direction, $\hat{\mu}_i^{(o)}$, resulting in the following relation between the time step $\de{t}$ and $R$
\equa{dmu_1}
 \de t = a \frac{R^2}{T^*} = a R_0^2   
\auqe
where $a$ is a constant and we have introduced $R_0$, independent of $T^*$. The constant $a$ is a function of both the gyromagnetic 
factor $\ga$ and the damping coefficient $\al$ and can be calculated, at least in principle once the value of $\al$ is known. 
Conversely, we consider that it includes the microscopic parameters, not 
explicitly taken into account in the effective one spin model and our alternative approach is to keep $a$ as a fitting parameter, 
instead of determining its value from known values of $\ga$ and $\al$. 
Notice that in Ref.~\cite{cheng_2006}, the relation~(\ref{dmu_1}) is deduced without starting from the equilibrium condition, 
and includes the value the acceptance rate $\Gamma$ of the MC process 
used with the heat-bath acceptance rate to the first order in $\gD{}E$ as
opposed  to the 
initial mapping performed in~\cite{nowak_2000}.
Therefore, at least in principle, equation~(\ref{dmu_1}) holds only for the heat-bath acceptance rate and for sufficiently 
small displacements so that the first order expansion of $\Gamma$ holds.

Now, from equation~(\ref{dmu_1}) we deduce that the time duration corresponding to $N_{MCS}$ Monte Carlo steps characterized 
by a constrained aperture cone $R^2=R_0^{2}T^*$ is $T^*$ independent and is given by 
 \equa{dt1}
   \gD t = a N_{MCS}R_0^2 
 \auqe
from which we introduce a unit of time for the TQMC scheme corresponding to $\gD{}N$ Monte Carlo steps
 \equa{dt2}
   u_t =  \gD{}N R_0^2/2.
 \auqe
This definition holds {\it a priori} only when the heat-bath acceptance rate is used and $R_0$ is small enough. 
The factor $1/2$ is introduced for convenience (see below equation~(\ref{dt3})).
In order to generalize to the Metropolis acceptance rate, 
we first notice the disymmetry of the latter with respect to $\gD{}E=0$ and second that for small displacements, where 
the heat-bath acceptance rate leads to $\Gamma\sim{0.5^-}$, the Metropolis one
leads to  $\Gamma$  being closer 
to $\Gamma\sim{1^-}$. 
Indeed if we assume that about half of the moment trials lead to $\gD{E}<0$ where $\Gamma_{met}=1$ then in average and still 
for small values of $\be\gD{E}$ the Metropolis acceptance rate is about $\Gamma_{met}\simeq(1-\be|\gD{E}|/2)$
 while $\Gamma_{hb}\simeq{}(1-\be\gD{}E/2)/2\sim{0.5}$, we can extend equation~(\ref{dt2}) to 
$u_{t,met}\simeq{2}u_{t,hb}$ 
when using the Metropolis acceptance rate instead of the heat-bath one. 
For not sufficiently small values of $R_0$, some deviations may occur (particularly for the Metropolis sampling).
In such cases we may extend equation~(\ref{dt2}) to $u_t=\Gamma_{acc}\gD{}N R_0^2$ where $\Gamma_{acc}$ is  the
value of the acceptance rate for a given temperature. For a MC path performed at fixed $T^*$ as a relaxation process,
$\Gamma_{acc}$ is obviously a constant. In our simulations, we limit ourselves to reasonably small values of $R_0$
for which $\Gamma_{acc}$ is very weakly dependent of $T^*$ in the range of temperatures considered in the MC 
study~\cite{note_gamma}. We thus propose to define an operational unit of time, both $T^*$ and sampling choice independent,
 \equa{dt3}
   u_t = \bar{\Gamma}_{acc}\gD{}N R_0^2
 \auqe
where $\bar{\Gamma}_{acc}$ is an average over the temperatures considered, excluding the high temperature range where 
the dynamics become reversible and the time unit is no longer relevant.
Notice that in the framework of the heat-bath acceptance rate and when $R_0\tend0$, equations~(\ref{dt2}) and~(\ref{dt3}) coincide.
A note should be made about the precise determination of $\bar{\Gamma}_{acc}$.
For isolated particles at $T^*=1$ and Metropolis scheme, $u_t$ is increased by a factor 3.4 when going from $R_0=0.2$ to 0.4. 
This contrasts with the variation of $\Gamma_{acc}$ in the significant range of temperature, 
excluding the reversible region: we get ca.~0.6\% and 1.3\% from $T^*=0.7$ to 1 at $R_0=0.2$ and 0.4 respectively. 
The error induced  in $u_t$ by the variation of $\Gamma_{acc}$ is thus negligible. The unit of time $u_t$ varies essentially through the 
simulation parameters, $\gD{}N$ and $R_0$.
\\ \indent
From this point, we have to distinguish between the instantaneous MC time step $i$, which is nothing but the index of 
Monte Carlo step (as usual one MC step corresponds to a trial move for all the moments) and the MC time $t_{MC}$, 
which includes the unit of time as defined by either equation~(\ref{dt2}) or~(\ref{dt3}), namely $t_{MC}=u_{t}.i$.
In most of our simulations, especially for relaxation process, we divide the whole number of Monte Carlo steps, say $N_m$,
into intervals of $\gD{}N$ playing the role of a finite $\gD{}t$ time interval. Then we consider sets of $\gD{}N$ steps 
which will be labelled by $i_{MC}\;\in[1, N_m/\gD{}N]$ and define the MC time
as $t_{MC}=i_{MC}u_t$. \\ \indent
This  approach is compatible with the one of Melenev~\cite{melenev_2012}, where instead of
the introduction of the unit of time,~(\ref{dt2}, \ref{dt3}), the number of MC
steps $N_{rel,T_0}$ corresponding to the relaxation time  $\tau_0$ for a given
temperature $T_0$  is used to define dimensionless time :
$t^*=i_{MC}/N_{rel,T_0}=t/\tau_0$. 
\cor{
Note that $t_{MC}$ and $t^*$ are dimensionless times. In the next section, we will explain how it is possible to map this 
simulation time with a time in seconds by using experimental data of specific systems.
} 
\\ \indent
The observables considered in this work are, besides the total energy, on the one hand the magnetization $M(t_{MC})$ 
projected on the direction of the external field $\hat{h}$
as the thermal average of the instantaneous $m(i)$ over $\gD{}N$ MC steps and
a large number (100 or 200) realizations of the systems as the studied
properties are ''time dependent'' :
\equa{M_t1}
 M(t_{MC}) = <m(i_{MC})>_{\gD{}N} = \frac{1}{\gD{}N}\sum_{i=i_{MC}\gD{}N}^{(i_{MC}+1)\gD{}N} m(i)
  \blanc \textrm{with} ~~ m(i) = \frac{1}{N}\sum_{k=1}^N \hat{\mu}_k\ldotp\hat{h}.
\auqe
On the other hand the susceptibility $\chi(\om)$ in terms of the frequency $\om$, 
determined from the method introduced by Andersson {\it et al.}~\cite{andersson_1997}. 
For the latter, the external reduced field is taken as $h(t)=h_0sin(\om{t})$ and we have
\equa{chi1}
        \chi'  = \frac{2}{h_0 N_{p}} \sum_{i=1}^{N_p} sin(\frac{2\pi}{N_p}i) m(i)
 \blanc \chi'' = \frac{2}{h_0 N_{p}} \sum_{i=1}^{N_p} cos(\frac{2\pi}{N_p}i) m(i)
\auqe
where $N_p$ is the number of MC steps corresponding to one period, {\it i.e.}
$N_p.u_t=2\pi/\om$. The period corresponds to the measuring time $t_m=N_p u_t$. On a practical 
point of view, the whole MC run involves $N_m$~MC steps covering a number $p$ of periods, 
the averages involved in equation~(\ref{chi1}) are performed over the last $p/2$ periods.
%
Equation~(\ref{chi1}) is written for $\gD{}N=1$ for the sake of simplicity, but can easily be translated to the more 
general case of $\gD{}N>1$ as 
\equa{chi2}
      \chi'  = \frac{2}{h_0 N'_{p}} \sum_{i=1}^{N'_p} sin(\frac{2\pi}{N'_p}i)<m(i\gD{}N)>_{\gD{}N}
      ~ \textrm{with} ~ N'_p = \frac{N_m}{\gD{}N}\frac{1}{p}
\auqe
and similarly for $\chi''$. In any case, when translated to the {\it actual} MC time, the value of $\gD{}N$ does not play any role. 
From the magnetization $m(i_{MC})$, we will mainly look at the relaxation either in the absence of an external field and starting
for instance from a state with all the moments in a fixed direction, or in the presence of a static external field $h_0$
with the system evolving to the equilibrium value $m_{eq}(h_0)$.

 \section {Results}
 \label {results}
The characterization of the time evolution of dynamic properties calculated in the framework of the TQMC is done first 
through the relaxation of the magnetization $M(t)$ (Unless indicated, in what follows $t$ stands for the MC time $t_{MC}$). 
For this purpose we consider the ensemble of NP
with all the easy axes either aligned in the $\hat{z}$ direction, textured case, or randomly distributed  
and with all the moments aligned along the
$\hat{z}$ axis with $\mu_z=1$ as the initial state. Then we let the system evolve in the absence of external field towards
the equilibrium state characterized by $M_z=0$ according to the standard Néel process.

\begin {figure}
     \includegraphics [width = 0.66\textwidth]{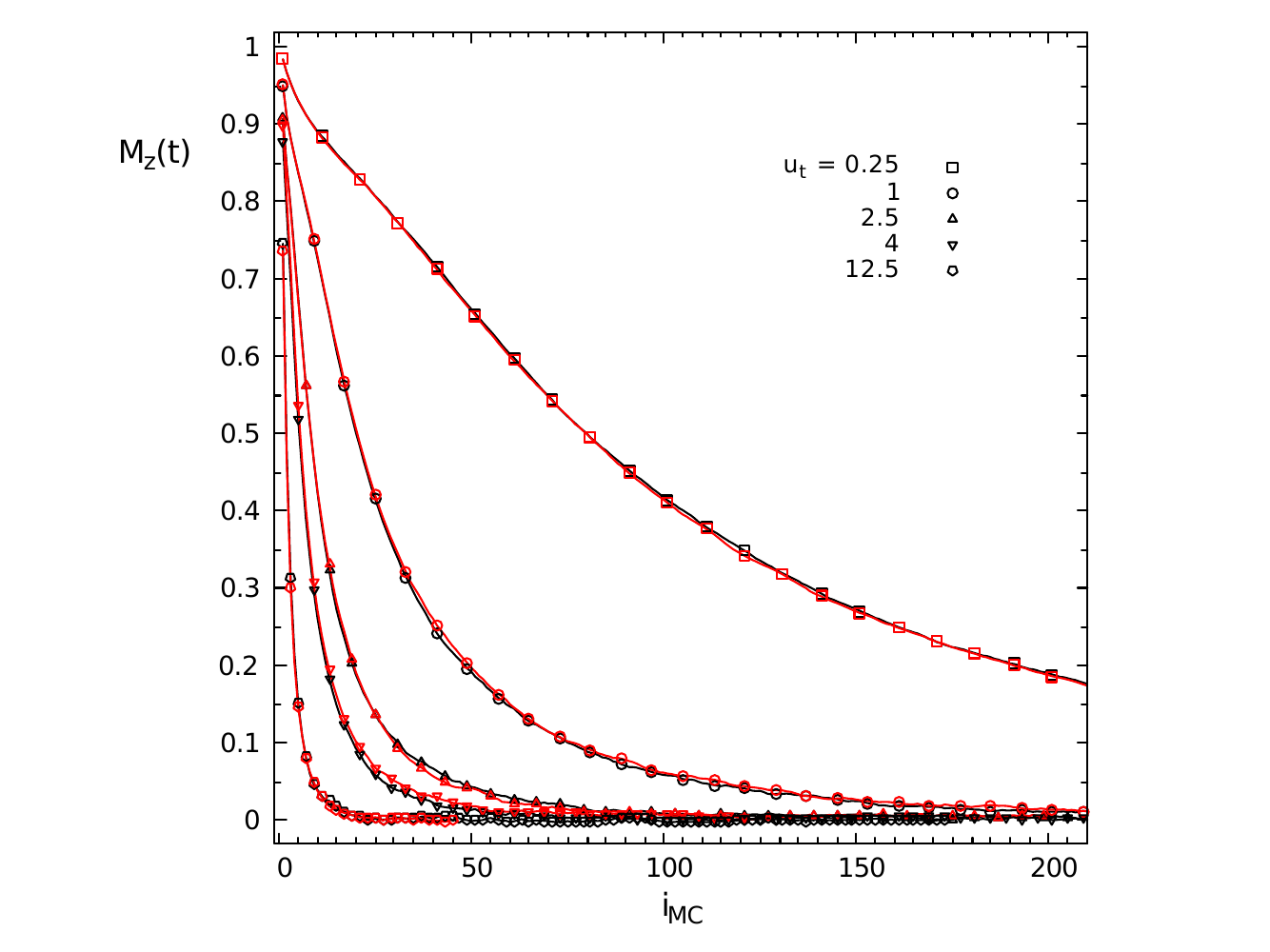}
   \caption {\label {fig1}
   Relaxation of the magnetization from $M_z=1$ in absence of external field. FCC lattice with 
   500 NP and easy axes distribution textured along $\hat{z}$, $\la_u=4$, $\la_d=1$ and $T^*=1$. 
   Two realizations for each values of $u_t$ (see equation~(\ref{dt2})) with $u_t$ going from 0.25 to 12.5 as indicated.
   These are obtained with pairs  ($\gD{}N,R_0$) = (200, 0.05) and (800, 0.025); (200, 0.10) and (3200, 0.025);
   (2000, 0.05) and (8000, 0.025); (200, 0.2) and (800, 0.10); (2500, 0.1) and (10000, 0.025)
   for $u_t$ = 0.25 to 12.5 respectively.
   }
  \end {figure}

\begin {figure}
     \hskip -0.19\textwidth
     \includegraphics [width = 0.66\textwidth]{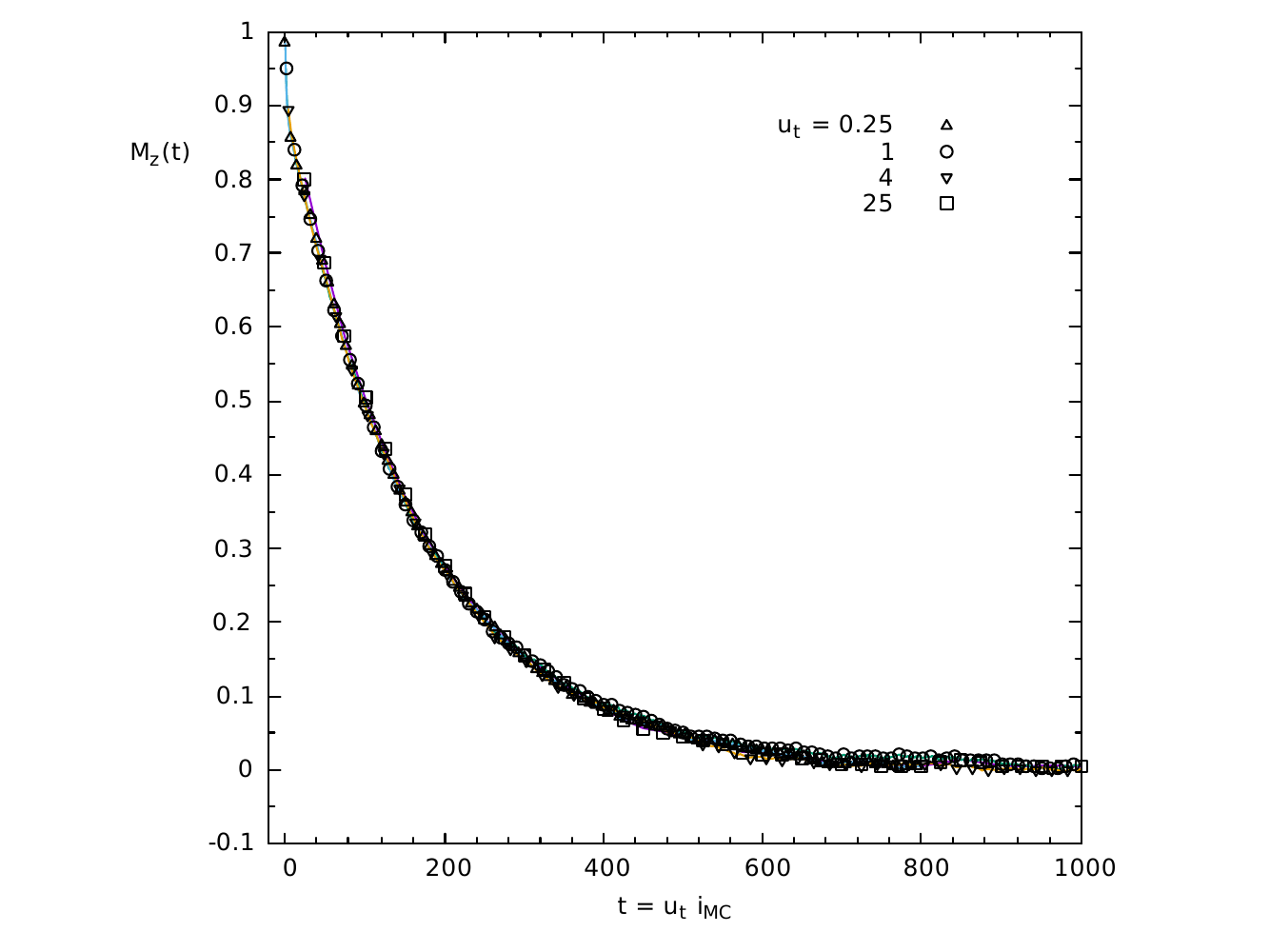}
     \hskip -0.14\textwidth
     \includegraphics [width = 0.66\textwidth]{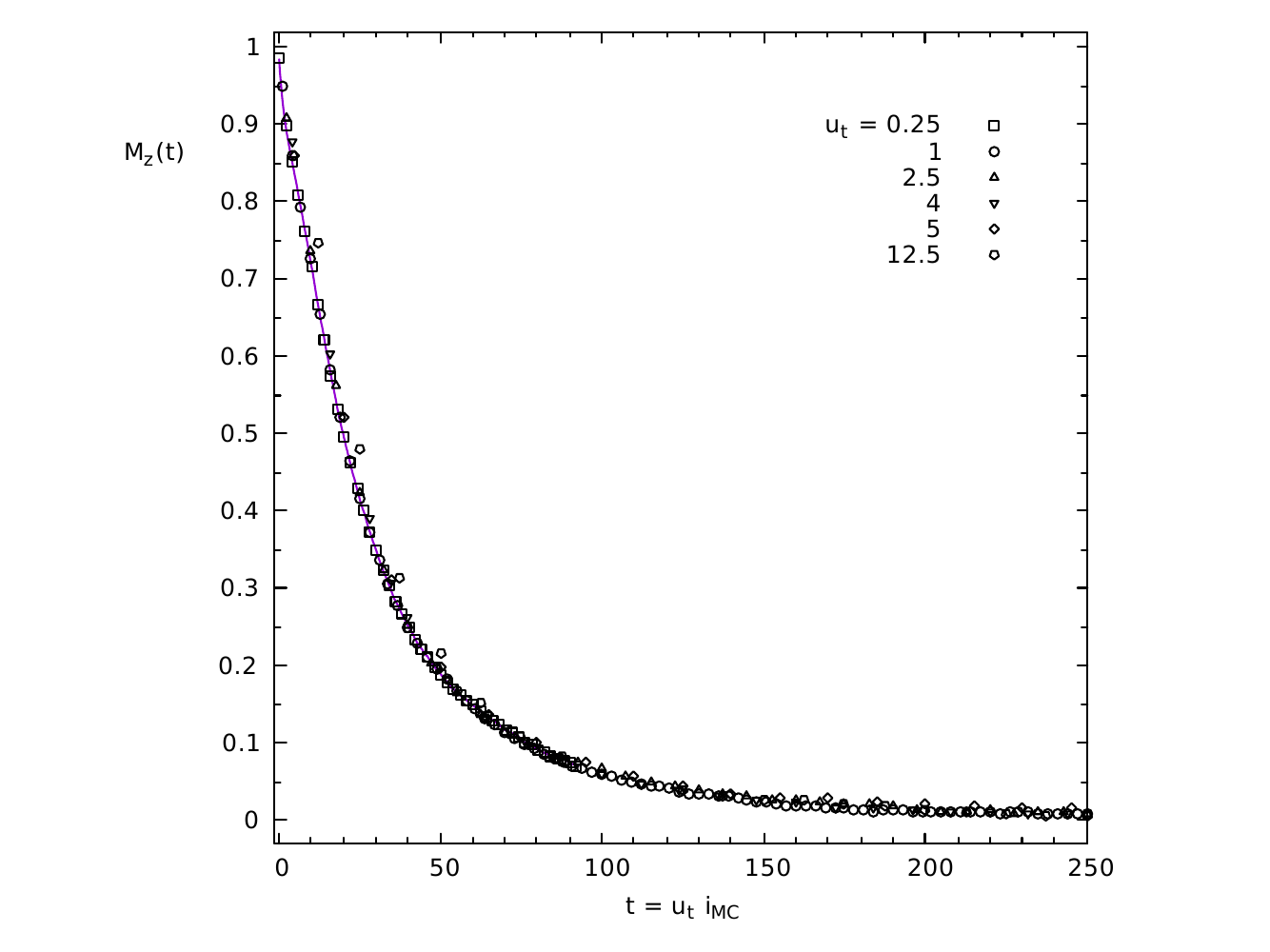}
   \caption {\label {fig2}
   Relaxation of the magnetization starting from $M_z=1$, displayed in terms of the MC time
   with $u_t$ from equation~(\ref{dt2}). Heat bath sampling.
   FCC lattice with 500 NP, easy axes distribution textured along $\hat{z}$ and $\la_u=4$.
   Left: non interacting system; 
   right: interacting system with $\la_d=1$. $T^*=1$
   }
  \end {figure}

\begin {figure}
    \hskip -0.20\textwidth
    \includegraphics [width = 0.66\textwidth]{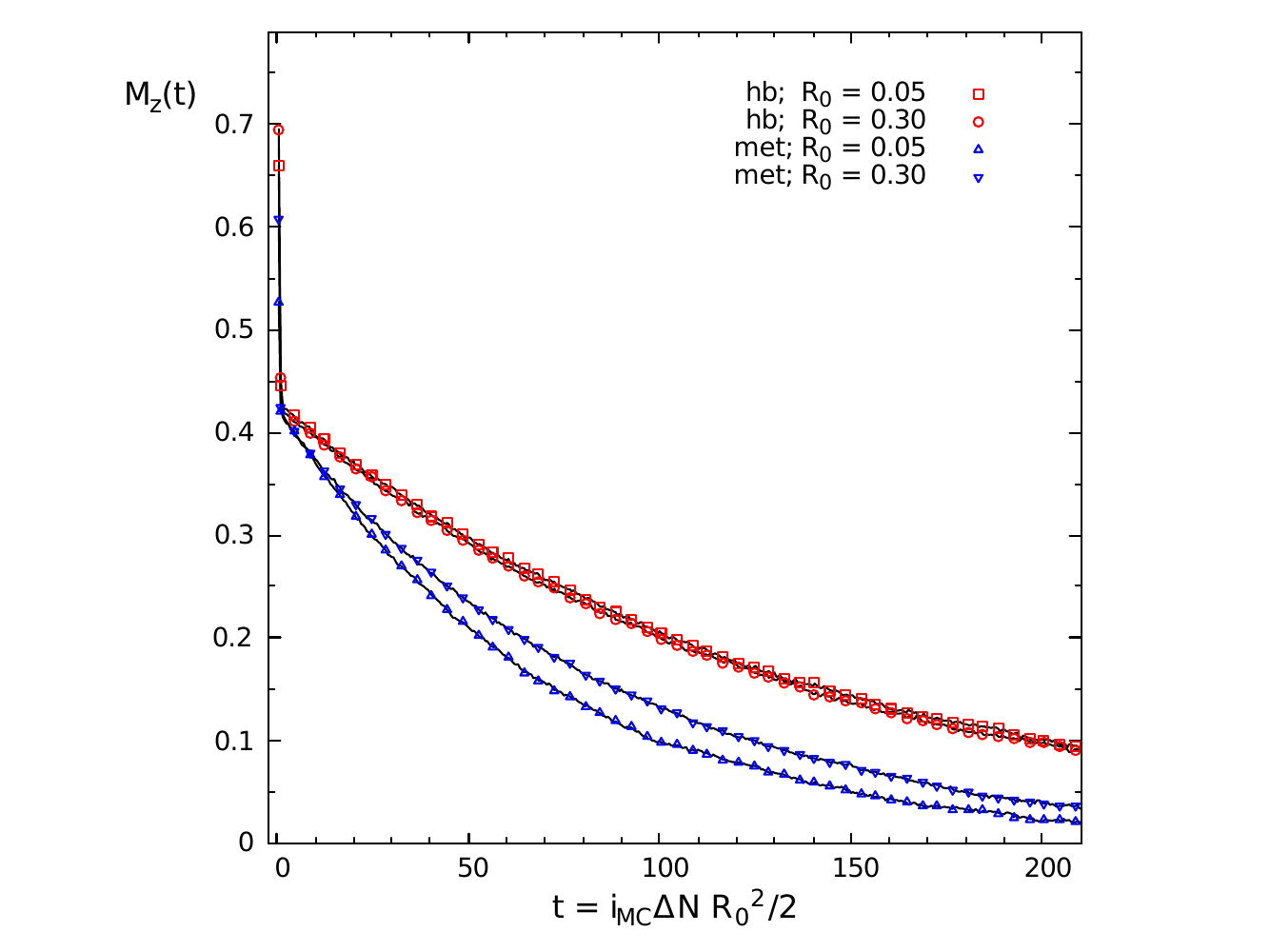}
    \hskip -0.15\textwidth
    \includegraphics [width = 0.66\textwidth]{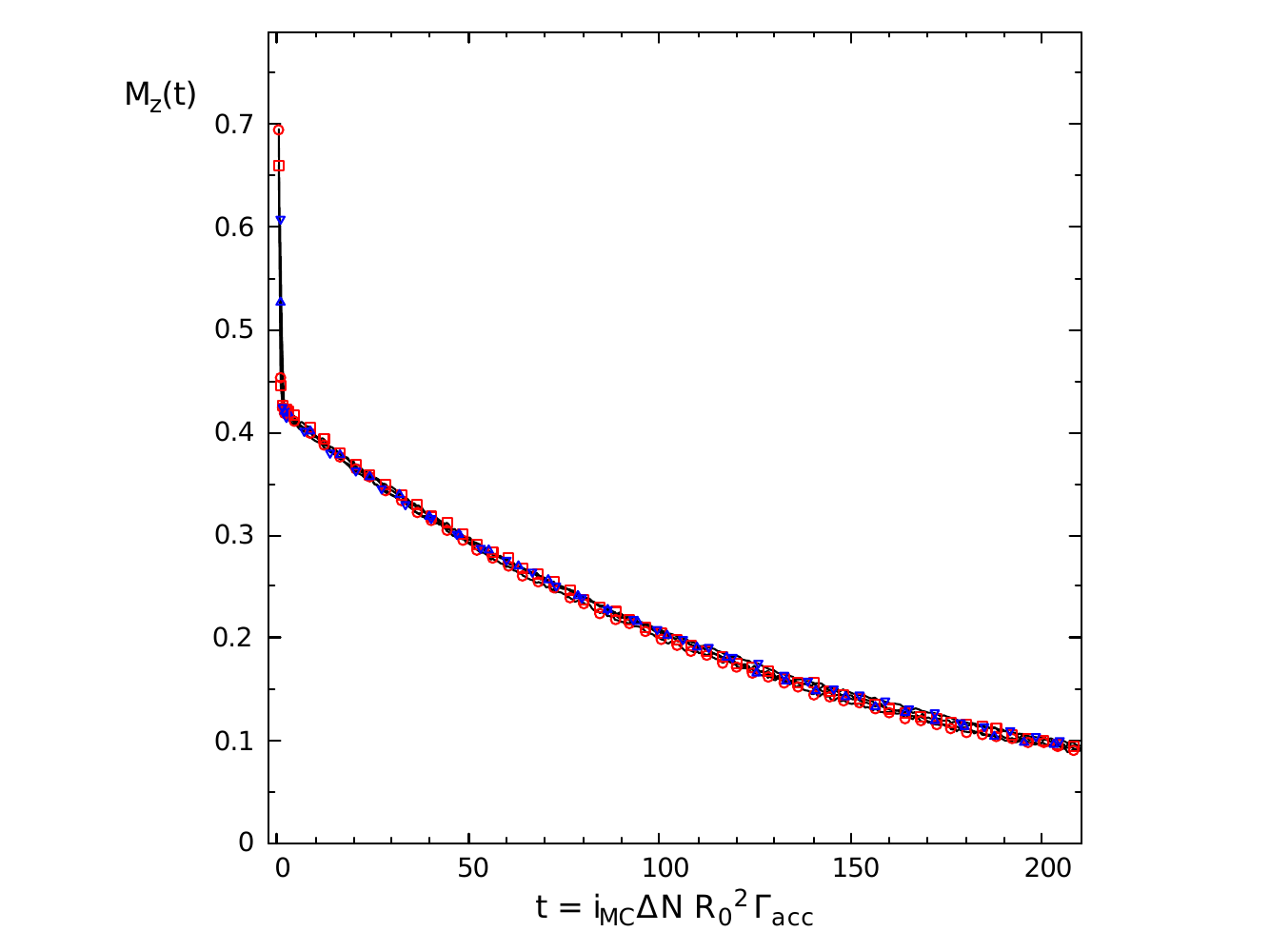}
   \caption {\label {fig3}
   Relaxation of the magnetization starting from $M_z=1$, 
   calculated from either the heat-bath (hb) or Metropolis (met) sampling acceptance rates and two values for $R_0$ 
   as indicated, without (left) or with (right) the extension of the 
   definition of the MC time unit when going from equations~(\ref{dt2}) to~(\ref{dt3}).
   Ensembles of non interacting particles with $N=500$, random distribution of easy axes and $\la_u=8$.
  }
  \end {figure}

\begin {figure}
  \includegraphics [width = 0.66\textwidth]{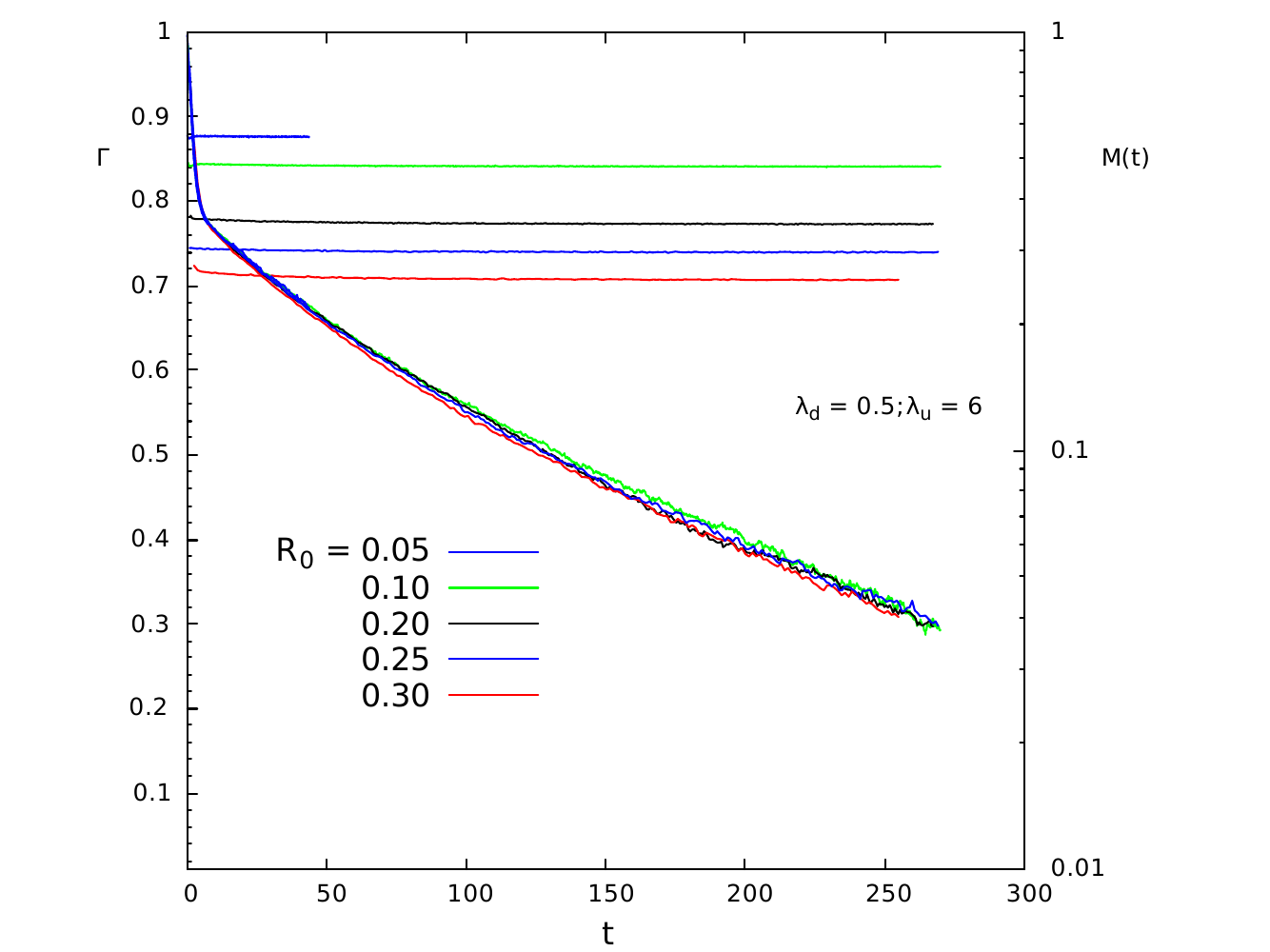}
  \caption {\label {relax_ddi_05_6}
  Relaxation of interacting NP with dipolar interactions for 5 different realizations of the unit of time $u_t$ 
  from equation~(\ref{dt3}) in the Metropolis sampling. $\la_d=0.5$, $\la_u=6$, $T^*=1$. 
  Flat curves on the upper part: (\cor{left} 
  y-axis) 
  evolution of the acceptance rate; 
  curves on the lower part (\cor{right} 
  y-axis) magnetization $M(t)$ starting form $M(t)=1$, in log scale.
  }
  \end {figure}

We first check that $u_t$ defined in equation~(\ref{dt2}) corresponding to the
heat-bath acceptance rate actually
plays the role of a unit of time for the TQMC Monte Carlo.
To this aim, we consider a set of relaxation runs characterized by the same value of $u_t$ obtained with different couples 
of values of ($\gD{}N_{MCS},R_0^2$) for a typical case of DDI interacting particles (fig.(\ref{fig1})). 
A very similar result is obtained for non interacting particles. As one sees, the relaxation curves corresponding to a 
given value of $u_t$ coincide. 
In the same way, we make coincide the $M(t)$ corresponding to different values of the 
Monte Carlo unit of time $u_t$ when displayed in terms of $t=i_{MC}u_t$. (fig.(\ref{fig2})). At this point only the heat-bath
acceptance rate for reasonable values of $R_0$ is used according to ~\cite{nowak_2000}. In order to test our suggestion allowing 
to use also the Metropolis acceptance rate according to the extension from equation~(\ref{dt2}) to~(\ref{dt3}) we perform 
magnetization relaxations by using both acceptance rates. See the left panel of figure~(\ref{fig3}). As expected the Metropolis 
algorithm is more efficient than the heat-bath one. However, once again we make the corresponding $M(t)$ coincide by using the 
unit of time given in equation~(\ref{dt3}), see the right panel of figure~(\ref{fig3}). 
Another example is given in figure~(\ref{relax_ddi_05_6}) in the case of NP interacting through DDI where we also include the 
variation of $\Gamma_{acc}$ along the whole relaxation path. The latter is clearly constant although the relaxation 
does not follow a single exponential decay due to the DDI.
A similar behavior has been mentioned in~\cite{zapata_2021} for non interacting MNP at least for $\Gamma_{acc}\geq{}0.40$.

We also note from figures~(\ref{fig3},\ref{relax_ddi_05_6}) that
one can extend the value taken by $R_0$ at least up to $\sim{}~0.3$ in order to satisfy the dynamics. 
It is noteworthy that the efficiency is proportional to $R_0^2$. As an illustration, in figure~(\ref{relax_ddi_05_6}). 
the paths corresponding to $R_0=0.05$ and 0.3 use 1000 and 200 steps of $\gD=20$ MCS respectively although 
the first one is $\sim$~6 times shorter in real time.

\begin {figure}
  \includegraphics [width = 0.66\textwidth]{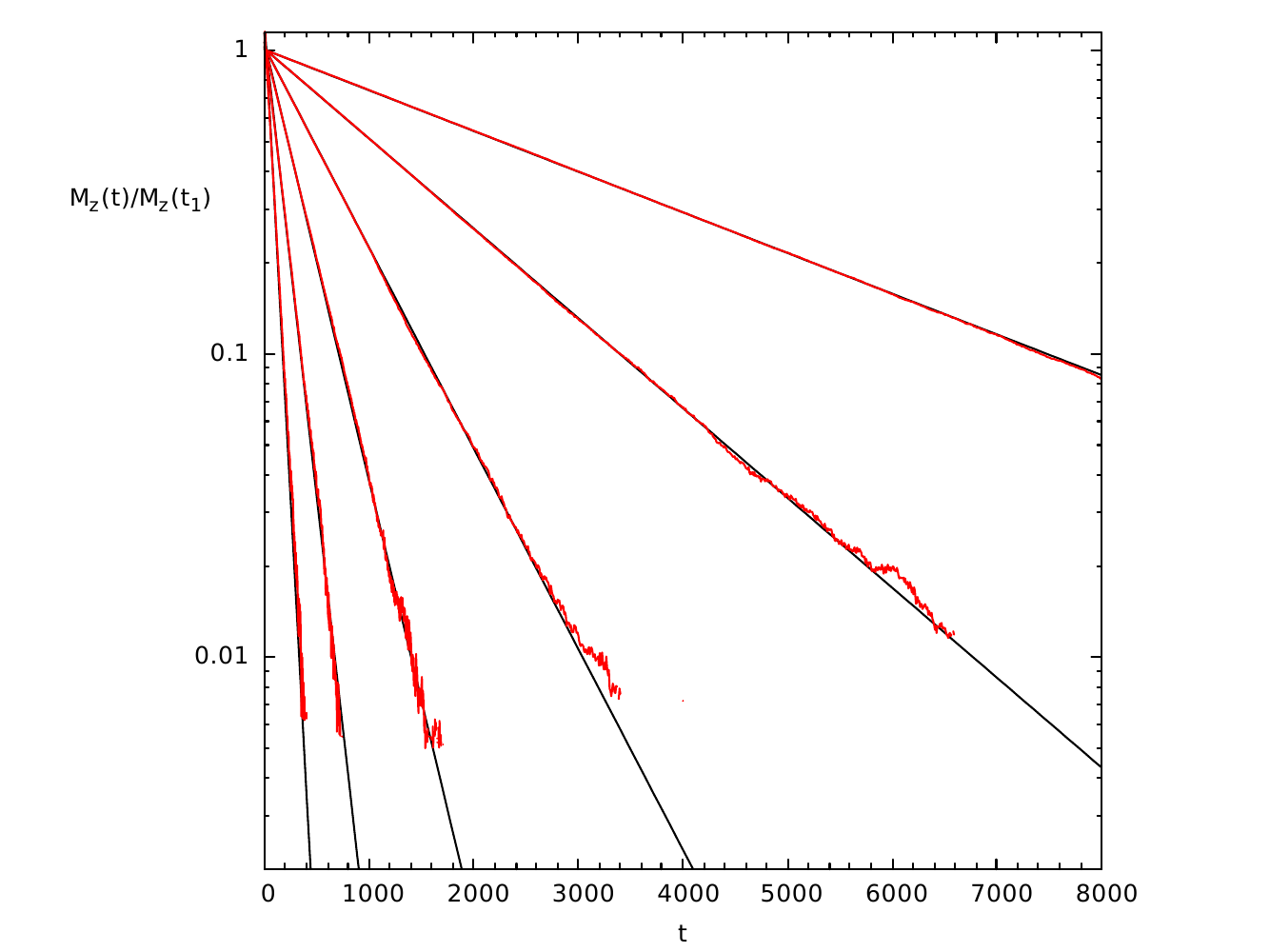}
  \caption {\label {relax_rand}
  Relaxation of $M_z(t)$ in terms of t (in MC time units) relative to the value of $M_z(t_1)$ for ensembles of 1720 non interacting NP
  at $T^*=1$ and $\la_u=10$, 9,8,7,6 and 5 from right to left. $t_1=19$.
  }
  \end {figure}

\begin {figure}
  \includegraphics [width = 0.66\textwidth]{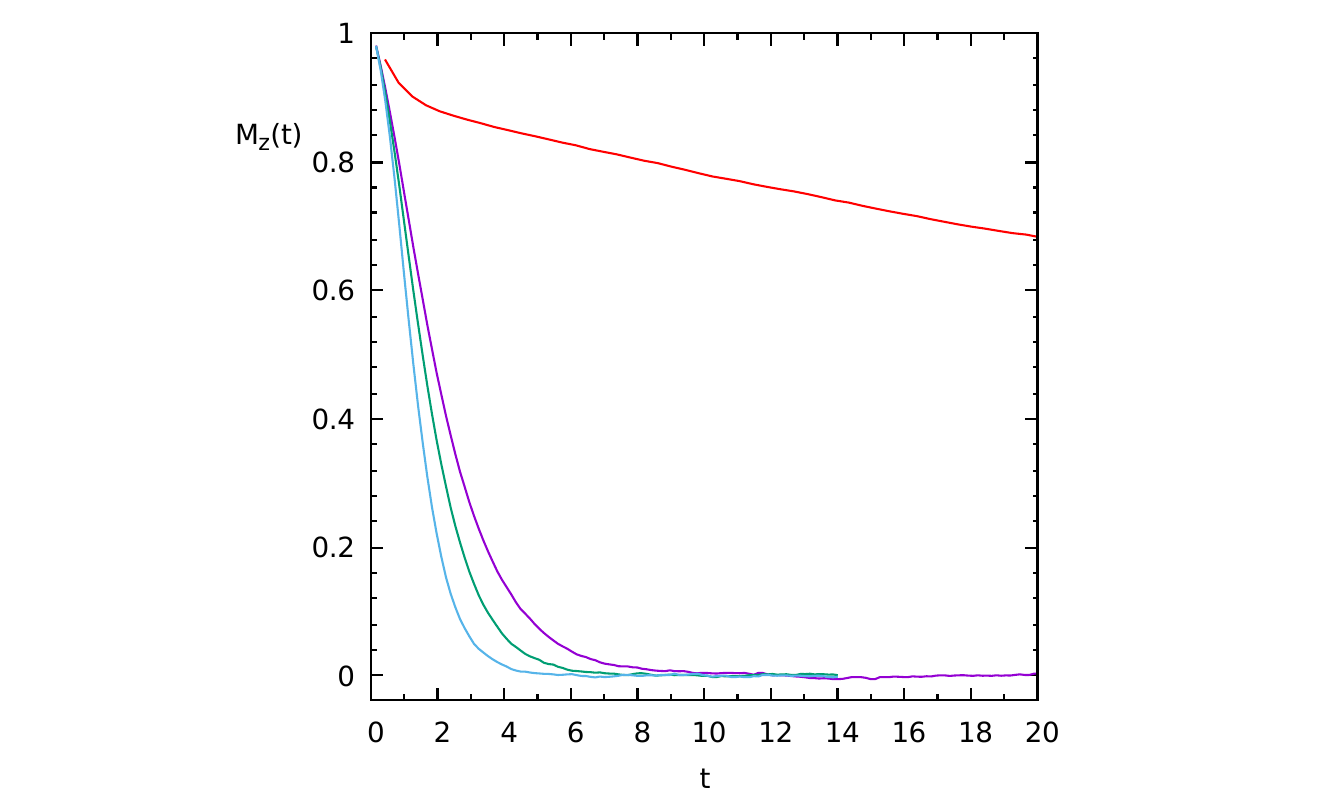}
  \caption {\label {relax_xz}
   Magnetization relaxation in absence of external field, starting from $M_z(t=0)=1$ for a non 
   interacting system with a textured distribution of easy axes along either the $\hat{z}$-axis 
   and $\la_u=5$ (top curve) or the $\hat{x}$-axis and $\la_u=5$, 7 and 10
   (rest of the curves from left to right) probing the 
   longitudinal $\tau_{\parallel}$ and transverse relaxation times $\tau_{\perp}$ respectively.
   $T^*=1.0$.
  }
  \end {figure}

In order to clearly identify the relaxation time, we let evolve $M(t)$ up to some value, say $t_1$
and look at the relaxation of $M(t)/M(t_1)$ in terms of $(t-t_1)$. 
Doing this, we exclude the very short time relaxation process. We thus obtain the relaxation time $\tau$ from the fit of 
$M(t)/M(t_1)$ on $\exp(-(t-t_1)/\tau)$. An example is shown in figure~(\ref{relax_rand}). 
It is worth mentioning  the longitudinal (inter-well) 
relaxation time, $\tau_{\parallel}$ is much larger than the transverse (intra-well) one, $\tau_{\perp}$ (see figure~(\ref{relax_xz})).
As a result, the relaxation time for a random distribution of easy axes is quite close to $\tau_{\parallel}$.

\begin {figure}
   \includegraphics [width = 0.66\textwidth]{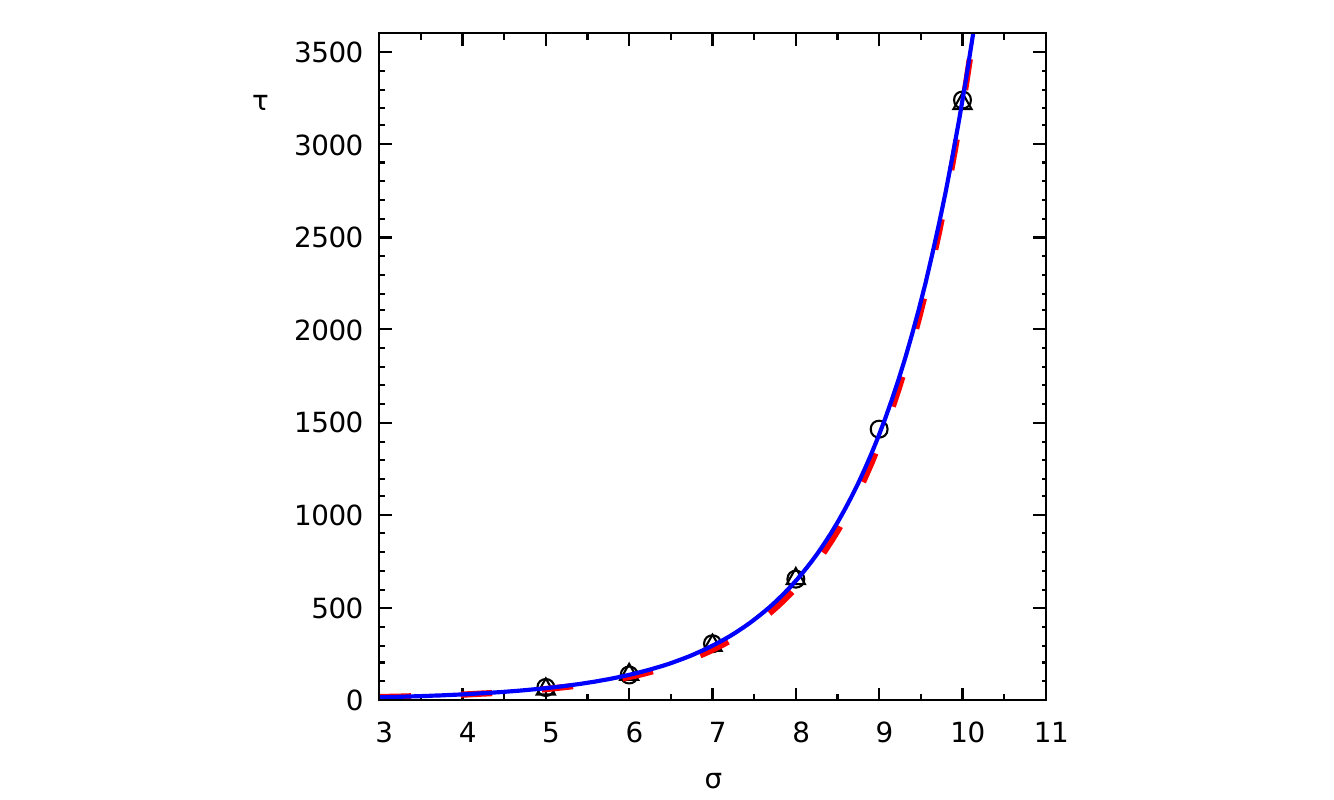}
   \caption {\label {brown_bessais}
   Relaxation time $\tau$ in terms of $\s$ at $T^*=1$ for non interacting NPM with random (circles) or textured along
   $\hat{z}$ (triangles) distribution of easy axes.
   $\tau$ is expressed in $u_t$ units from equation~(\ref{dt3}).
   Solid and dashed lines are the fits on the analytical results taken from 
   Refs.~\cite{bessais_1992} and ~\cite{brown_1963} respectively.
   }
  \end {figure}

\begin {figure}
    \hskip -0.20\textwidth
    \includegraphics [width = 0.66\textwidth]{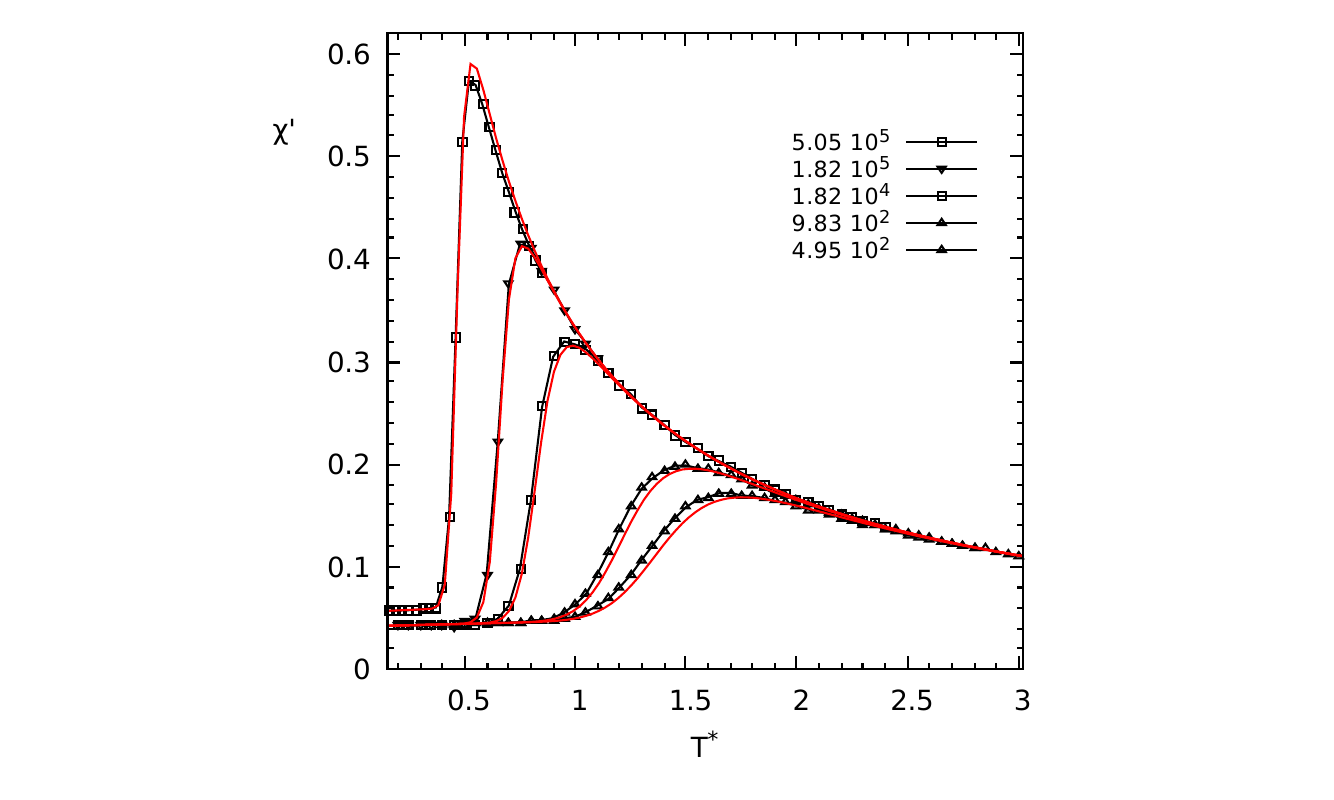}
    \hskip -0.15\textwidth
    \includegraphics [width = 0.66\textwidth]{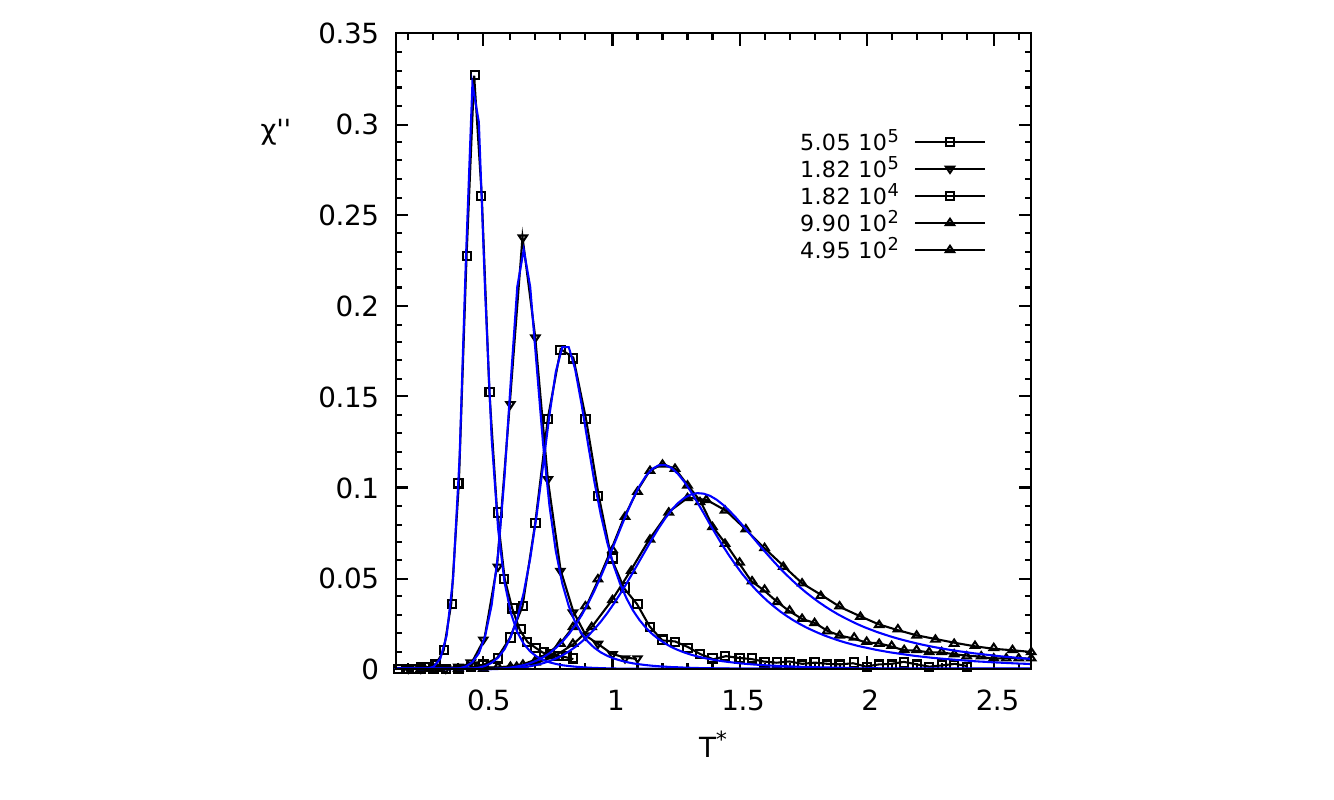}
    \caption {\label {test_sh_ed0}
    Comparison between the susceptibility calculated from the approximation of~~\cite{shliomis_1993} 
    with $\tau$ taken from~\cite{bessais_1992} (solid lines)
    and our TQMC simulations (symbols) in the absence of DDI and with random distribution of the easy axes. 
    Left (right) panel displays  curves 
    for $\chi^{'}$ ($\chi^{''}$). The curves are indexed by the actual MC measuring time (see text). 
    In each panel, the left hand side curve ($t_m=5.05\;10^5$) is for $\la_u=6$ and the other ones for $\la_u=8$.
    The product $\chi^{'}_{max}.h_0$ ranges from 0.048 to 0.17 and is seen as a criterion for the LRT regime to be valid 
    ($\chi^{'}_{max}.h_0\;\lesssim{}0.15$).
    }
  \end {figure}

\begin {figure}
    \hskip -0.20\textwidth
    \includegraphics [width = 0.65\textwidth]{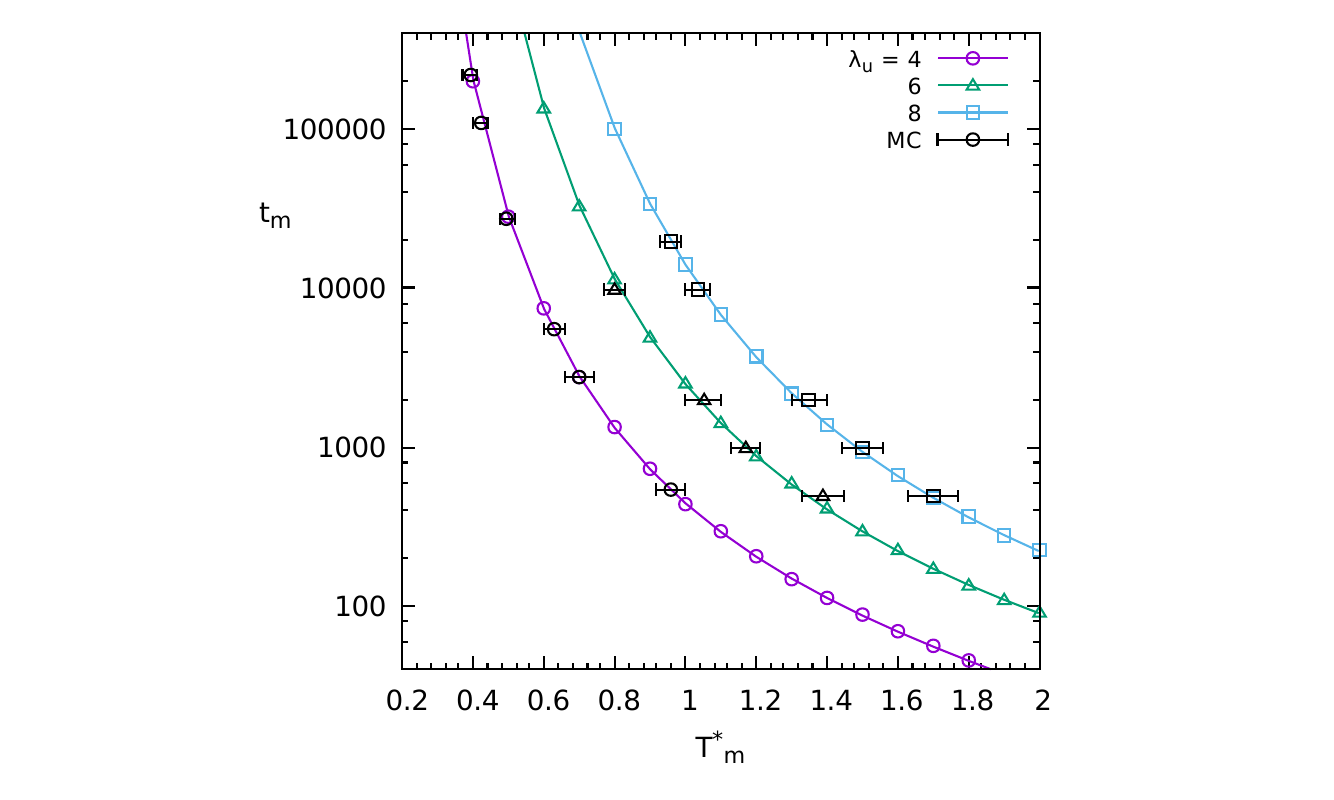}
    \caption {\label {tm_tmax_X}
    MC measuring time $t_m$ in terms of the temperature $T^*_m$ of the maximum of $\chi^{'}(\om=2\pi/t_m)$. 
   Colored lines/symbols stand from results obtained from the approximation
   of~~\cite{shliomis_1993}. Black symbols stand for TQMC results.
   Circles, triangles and squares are for $\lambda_y=4,6,8$ respectively. Non interacting system with random distribution of easy axes. 
    }
    \end {figure}

In order to determine how the longitudinal relaxation time depends on $\la_u$ and $T^*$ for non interacting MNP, 
two sets of relaxation simulation runs are performed. First, we consider a set of values of $\la_u$ at constant $T^*=1$, which
remains to examine different values of $\s=\la_u/T^*$ at constant $T^*$ and second, a set of values of $T^*$ with the same value of 
$\s$, namely with $\la_u=T^*\s$. From the latter, we clearly evidence that $\tau$ behaves as $\tau=\exp(\s)f(\s)/T^*$ in
agreement with Refs.~\cite{brown_1963,bessais_1992}. 
For the function $f(\s)$ our simulations fit well the analytical results of either Brown~\cite{brown_1963} or 
Bessais {\it et al.}~\cite{bessais_1992}, 
both valid for 
\cor{$\s\ge{}2$ (here we consider only $\s>2.6$),} 
as is shown on figure~(\ref{brown_bessais}), namely 
\equa{tau_br_bes}
 \tau = b_0\frac{\exp(\s)}{\s^{3/2}T^*} \blanc \textrm{or} \blanc \tau = b_1\frac{\exp(\s)}{(1+\s/4)^{5/2}T^*}
\auqe
respectively.
We must notice that in the range of values of $\s$ considered here, we cannot distinguish clearly between these two analytical
forms of $\tau(\s,T^*)$.  
The relaxation time $\tau$ given in equation~(\ref{tau_br_bes}) is still unitless until an actual value in {\it second} is
associated to the constant $b_0$ or $b_1$. This is not straightforward, as already noted in~\cite{bessais_1992} for instance,
and is to be compared to the estimation of $\tau_0\sim{}10^{-8}-10^{-10}s$ generally used for the Arrhenius form. 
Following~\cite{brown_1963} as well as~\cite{bessais_1992}, the constants $b_0$ or $b_1$ should depend on the NP properties 
as $\sim\be_{0}vM_s\al/\ga$ where $\al$ and $\ga$ are the damping and the gyromagnetic factors respectively. 
In any case such a correspondence would need the precise knowledge of the NP properties. 
In principle this could be achieved from the knowledge of $K_u$ and the experimental relaxation time, but from our
knowledge the latter is not available.
We consider that a better way to map the actual time to the MC time is
to focus on the behavior of $T_{max}(\chi')$ or $T_{max}(\chi'')$ in terms of $\om$ where $T_{max}(\chi')$ ($T_{max}(\chi'')$)
is the temperature of the maximum of the AC susceptibility $\chi'$ ($\chi''$).
\\
Now in the case of non interacting NP, we compare our simulation results for the AC susceptibility $\chi{}(\om)$ 
to the analytical result obtained from the Shliomis and Stepanov model~\cite{shliomis_1993,garcia-palacios_2000} 
in terms of the reduced variable $y=\om\tau$ where only the longitudinal relaxation time is retained and 
$\tau=\tau_{\parallel}$, given for a random distribution of easy axes by
\equa{shliomis}
   \chi' &=& \frac{\chi_0 + \chi_{1}y^2}{1+y^2} \blanc \chi'' = \frac{y\chi_{\parallel}}{1+y^2} \\ \nonumber
   \textrm{with} &~& ~ \chi_0 = \frac{1}{3T^*} ~;~ 
   \chi_{1} = \chi_0(\frac{1}{\s} + \frac{1}{2\s^2} + \frac{5}{4\s^3}) ~;~ 
   \chi_{\parallel} = \chi_0(1 - \frac{1}{\s} - \frac{1}{2\s^2} - \frac{5}{4\s^3})
\auqe
where the limit $\s>>1$, which is valid down to $\s\simeq{}3.5$~\cite{garcia-palacios_2000}, has been taken.
In order to compare our simulated susceptibility with equation~(\ref{shliomis}), 
$y$ is taken with $\tau$ from equation~(\ref{tau_br_bes}) according to Ref.~\cite{bessais_1992} 
with the constant $b_1$ fitted from our TQMC simulations of the relaxation described above 
and $\om=2\pi(N_{p}u_t)^{-1}$ with $u_t$ given by equation~(\ref{dt3}). 
The agreement that we get between our simulated $\chi$ and the analytic results Ref.~\cite{shliomis_1993,garcia-palacios_2000}, 
give us confidence on the way we introduce the time unit 
\cor{$u_t$} 
since the latter involves 3 decades of frequency.
Notice that the nice agreement we get between our simulations and equation~(\ref{shliomis}) (see figure~(\ref{test_sh_ed0}))
holds only for sufficiently small values of the field amplitude $h$ in order to avoid non linear terms in the 
response to the AC field. As an heuristic criterion, we get from the maximum value of $\chi^{'}$, 
$h\chi^{'}_{max}\lesssim{}0.15$.   
\\
From $\chi'(\om,T^*)$ and $\chi''(\om,T^*)$ we can deduce the 
behavior of  $T'_{max}(t_m)$ and $T''_{max}(t_m)$ which are the locations of the maxima of $\chi'$ and $\chi''$ 
in terms of $T^*$ at fixed values of the measuring time $t_m=2\pi/\om$.  
On figure~(\ref{tm_tmax_X}), we give $T'_{max}(t_m)$ and $T''_{max}(t_m)$ resulting from both the analytical 
result of equation~(\ref{shliomis})~\cite{shliomis_1993} and the MC simulations in the case of the random 
distribution of easy axes. The curves displayed in figure~(\ref{tm_tmax_X}) show the dependence of $T^*_{max}$ on 
the measuring time $t_m$ for several values of $\la_u$. Note the time, temperature and energies remain expressed 
in terms of $u_t$ and $T_0$.
\\ \indent
To map the TQMC time on a true time we need to compare these curves to the equivalent ones obtained from real 
AC susceptibility experiments for a sample of non-interacting MNP with given anisotropy $K_{u}v(d)$. 
In order to do this, we first note that once a reference temperature $T_0$ is chosen, $T^*$ is translated into a real one 
and $\la_u$ gives $K_{u}v(d)$. Then from the $T_{max}^{'exp}$  found in experiments for a given $\om^{exp}$ 
\cor{measured in $s^{-1}$} 
we may find  the corresponding MC frequency $2\pi/(N_{p}u_t)$ for which $T_{max}^{'*}$ coincides with $T_{max}^{'exp}$ 
for the appropriate value of $\la_u$. 
\cor{This allow to quantify the TQMC time unit in such a way that $u_t$ represents a dimensioned time according to :} 
\equa{quant_1}
  u_t = \frac{2\pi}{\om^{exp} N_p}
\auqe
We emphasize that this mapping of $u_t$ is dependent of the specific system as it should be for the resulting mapping
of either $b_0$ or $b_1$ since the relaxation time is obviously system dependent.

To test our mapping procedure on an actual experimental system is not straightforward since the AC measurements
on truly isolated MNP assemblies are only few~\cite{de-toro_2013b,garcia-acevedo_2023}. 
In Refs.~\cite{de-toro_2013b,garcia-acevedo_2023} the DDI are made tunable through a non magnetic SiO$_2$ layer
of variable thickness.
The usual criterion to quantify the importance of the interactions between MNP from AC measurements is the value 
taken by the so-called Mydosh coefficient, namely the slope $\Phi=-\gD{}T_m(\om)/(T_m(\om)\gD{}\ln(\om))$,
($\simeq\;-\partial\ln(T_m(\om))/\partial\ln(\om)$ if the variation of $T_m$ remains weak)
with $\Phi\gtrsim{}0.1$ or noticeably smaller in the non interacting or interacting cases respectivelly.

As an illustration, we look at the experimental result of Ref.~[\cite{de-toro_2013b}] where $T_m(\chi{'})$ and 
$T_m(\chi{''})$ are reported for a likely non interacting MNP assembly thanks to the SiO$_2$ coating of the 
particles, when the thickness of the later is of 17\;$nm$ for maghemite NP of 8\;$nm$ in diameter. 
Using the anisotropy constant $K_u=3.0\;10^4Jm^{-3}$ given in [~\cite{de-toro_2013b}] and choosing $\la_u=6$ 
we get $T_0=97K$.

We 
\cor{quantify} 
$u_t$ from equation~(\ref{quant_1}) for $f=600\;Hz$: with $T_0$ this leads to map $T_m^{'}\simeq{}52\;K$ or
$T_m^{*'}=0.535$ which is done both by MC and the Shliomis-Stepanov~\cite{shliomis_1993} approximation, 
which is found to fit perfectly well the value of $T_m^{*'}$ obtained by the simulation, see figure~(\ref{test_sh_ed0}). 
We then get for the MC measuring time $N_p.u_t=5.05\;10^5$. Finally, we translate to the 
\cor{MC} 
measuring time corresponding to 10\;Hz 
and deduce the temperature $T_m^{*'}(\chi^{'}(f=10Hz)=0.385$ ($T_m^{'}=37K$) which is reasonably close to the 
experimental one ($T_m^{'}\simeq{}44\;K$). 
 
Similarly, from susceptibility curves given in Ref.~\cite{garcia-acevedo_2023}, we estimate the values of 
$T_m^{'}$ for the NP of diameter $d=15\;nm$ at $ca$ 185\;K and 155\;K for $f=10^3$ and 10\;Hz respectively.
Using $k^{eff}_{DC}\simeq{}3\;10^4Jm^{-3}$ from~\cite{garcia-acevedo_2023}, we get $T_0=318\;K$ with
the choice $\la_u{}=12$, leading to $T^*{'}_m=0.58$ for $f=10^3\;Hz$. 
We then obtain the MC measuring time $N_p.u_t=2.30\;10^{9}$ to map this reduced temperature value and from our method, 
where the approximation of equation~(\ref{shliomis}) is used, we obtain $T^*{'}_m=0.472$ to represent the temperature
peak of $\chi^{'}$ at 10\;$Hz$, leading to $T_m^{'}=150\;K$ close to the experimental value ($\sim{}155\;K$).
 
However, the comparison must be taken with care since $\Phi$ still takes a rather small value ($\sim$ 0.04) 
for the samples of Ref.~\cite{de-toro_2013b} and 
we thus cannot totally rule out a very weak interaction. Similarly, in Ref.~\cite{garcia-acevedo_2023}, $\Phi$ 
remains smaller than $\sim\;0.08$ and the AC blocking temperature $T_b(\om)$ as determined from the peak of $\chi^{''}$
depends on the SiO$_2$ layer thickness $\gD$ up to $\gD\simeq{}d$ where $d$ is the NP diameter.
Moreover, the comparison between the experimental susceptibility and either the MC simulated one or the 
result given by~\cite{shliomis_1993} should include the polydispersity even of reduced range, through a 
distribution of $\la_u$, expected to broaden the peak. 

Now we demonstrate how the quantification of time made with isolated particles holds when
the coupling between particles is not negligible with respect to the MAE.  

\begin {figure}
    \hskip -0.20\textwidth
    \includegraphics [width = 0.73\textwidth]{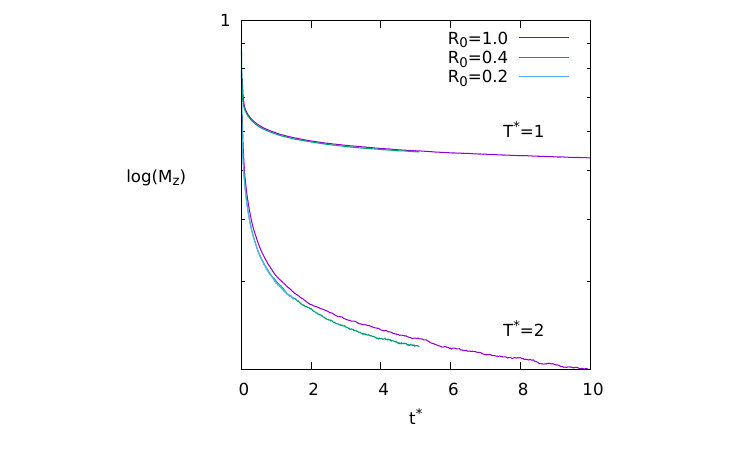}
    \hskip -0.10\textwidth
    \includegraphics [width = 0.53\textwidth]{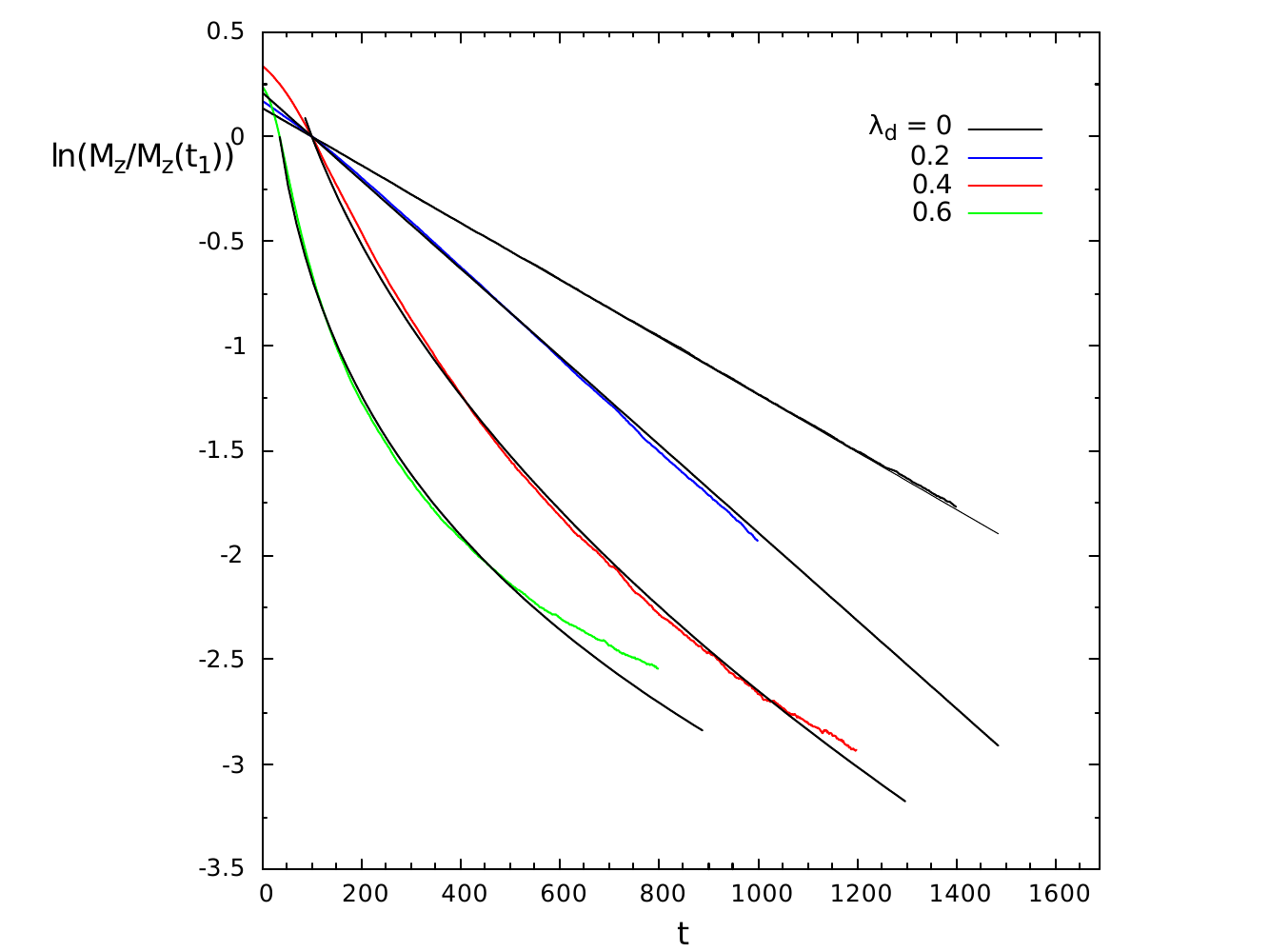}
    \caption {\label {JG_dyn}
    Magnetization relaxation for ensembles of interacting particles. 
    Left : short range Heisenberg interactions for a BCC lattice of 2000 NP
    with randomly distributed easy axes. $J=1$ and $\lambda_u=6.$ Two reduced temperatures $T_0^*=1$ (upper curves) and $T^*=2$
    (lower curves) and different dynamics with $R_0=1, 0.4$ and $0.2$. 
     Time is given as $t^*=i/N_{rel,T_0}$. \newline
     Right : long range DDI for a FCC lattice of 864 NP with easy axes distribution textured 
     along $\hat{z}$, $T^*=0.6$ and $\la_u=6$.
     and different values of $\la_d$ as indicated. Time is in $u_t$ units. The black thin lines are fits
     according to the stretched exponential :
     $M_z(t>t_1) \propto exp(-ct^{\be})$. $\be=0$ for $\la_d=0$ and 0.2, $\be=0.485,0.258$ for
     $\la_d=0.4$ and 0.6 respectively.  
    }
  \end {figure}

Fig~(\ref{JG_dyn}) show that starting from an ordered configuration of the dipoles along the $z-axis$, 
the relaxation of the system at different temperatures is no longer exponential for both 
a short range coupling or a long range DDI. Nevertheless, it is possible to use the properties of isolated
particles to define a 
\cor{unit of time} 
for the simulations of the interacting system.
Taking as a 
\cor{unit of time} 
the relaxation time $N_{rel,T_0}$ of the isolated particle with $R_0$ at $T^*_0$, it is possible 
to obtain the same relaxation curves using different dynamics ($i.e.$ Metropolis sampling with different values of $R_0$) but 
preserving the dependence in terms of temperature. In Fig~(\ref{JG_dyn}\;left) the relaxation is plotted as a function of a 
reduced time $t^*=i_{MC}/N_{rel,T_0}$. 
A similar plot should be obtained using the unit of time $u_t=R_0^2.\bar{\Gamma}_{acc}$. 
Note that the dominant parameter in the unit of time is $R_0^2.$ We checked that
with Metropolis algorithm and moderate $R_0$, the same results are obtained taking $\Gamma_{acc}$ 
either of the simulation for an isolated particle with $R_0^2$ or of one for the interacting particles.

\begin {figure}
    \hskip -0.20\textwidth
    \includegraphics [width = 0.65\textwidth]{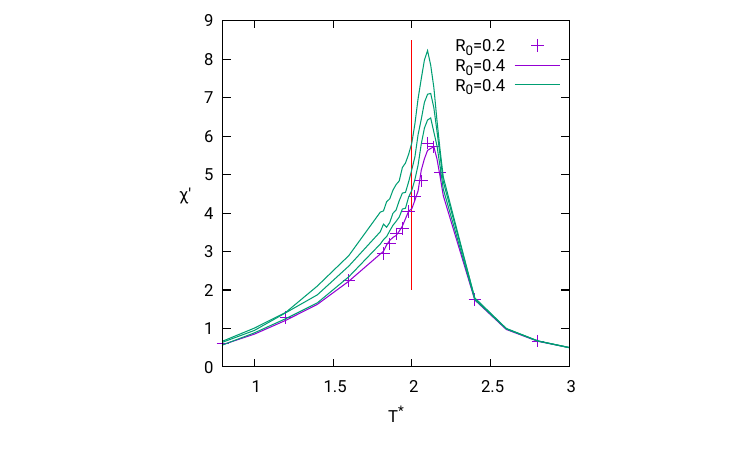}
    \caption {\label {JG_dyn_chi}
    In phase susceptibility in terms of temperature $T^*$ for interacting NP
    (same model as in Fig.~(\ref{JG_dyn}-left)) obtained from TQMC simulations for several
    values of the  dimensionless measuring times $t^*_m=N_p/N_{rel,T_0}.$ 
    Curve displayed with continuous lines are from bottom to top  for 
    $t^*_m=40.305, 48.366, 60.450$ and $80.610$  using $R_0=0.4.$
    Crosses and purple line compare AC susceptibility for
     $t^*_m=40.305$ obtained from dynamics with  $R_0=0.2$ and $R_0=0.4$ respectively. 
    The vertical line corresponds
    to the SPM/SSG transition temperature obtained from equilibrium MC simulations.}
    \end {figure}

\begin {figure}
    \hskip -0.00\textwidth
    \includegraphics [width = 0.65\textwidth]{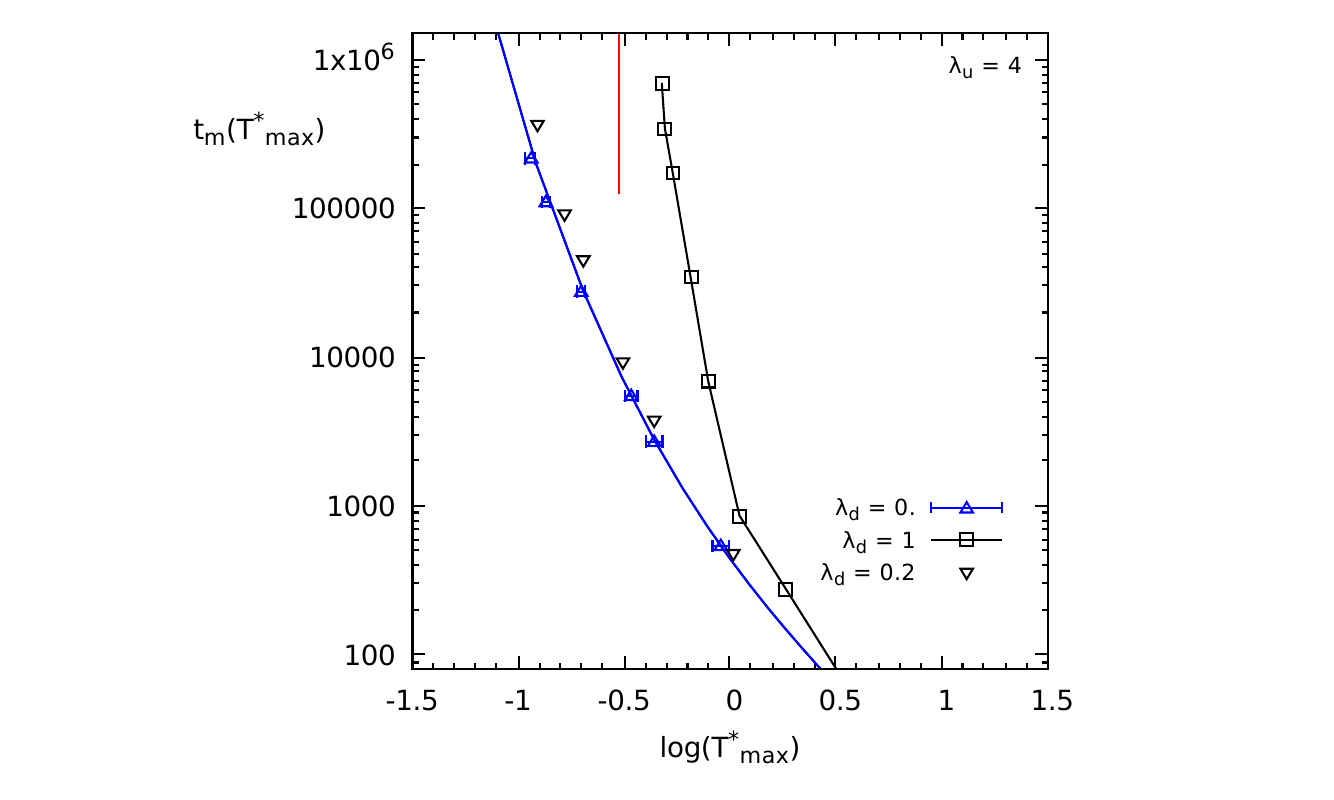}
    \caption {\label {x1_ed}
    MC measuring time $t_m$ in terms of the temperature $T^*_{max}$ of the maximum of $\chi^{'}(\om=2\pi/t_m)$
    for interacting NP with DDI, for $\lambda_u=4$ and the indicated values of
    $\lambda_d$ compared to the non interacting case. Data are obtained from TQMC simulations. 
    The blue line is the result from~\cite{shliomis_1993}. 
    The vertical red bar indicates the SPM/SSG transition temperature obtained from the fit of 
    equation~(\ref{freez1}). 
    }
  \end {figure}

This holds also for $\chi'(\omega)$ as shown 
in fig~(\ref{JG_dyn_chi}) where the same curves where obtained with different dynamical processes by adjusting $N_p$ with respect  
to $N_{rel,T_0}$. This allows to save computational time because short frequencies should be studied by using quick relaxing 
dynamics (with large $R_0$). Note however that only moderate values of $R_0$ should be considered and consistency tests should be performed. 
As an example, in Fig~(\ref{JG_dyn}\;left), for $T^*=2$ the same relaxation is obtained for $R_0=0.2$ and $R_0=0.4$ but not for $R_0=1$. 
\\~\\ \indent
Clearly in the presence of non negligible long range DDI, we get a relaxation
according to a stretched exponential behavior
as already found in~\cite{petracic_2006} (see figure~(\ref{JG_dyn}~right)).
Besides this relaxation process in the absence of external field, we have also studied the DDI and Heisenberg interaction
on the dynamics of $\chi^{'}$ and $\chi^{''}$. 
As it is well known the interactions between NP strongly reduce the 
decrease of $T_m(\chi^{'})$ as the measuring time $t_m$ is increased. 
This is the consequence of the collective behavior expected at low temperature leading to a freezing of the system 
with a power law dependence
\equa{freez1}
t_m \propto (T_m^{'} - T_c)^{-z\nu}
\auqe
with $T_c$ the super-paramagnetic (SPM)/super spin-glass (SSG) 
transition temperature between the SPM and the expected SSG phases. The decrease of 
$T_m^{'}$ together with the power law dependence of $t_m$ according to equation~(\ref{freez1}) are 
well reproduced in our simulations (see fig.~(\ref{x1_ed})).
\section {Concluding remarks}
\label {conclu}
The purpose of this work is mainly to revisit the TQMC scheme in order to
provide a way to map the simulation scheme on an actual time
scale. This is done first by showing that the precise control of the temperature dependent cone amplitude of the
moments displacements leads to the definition of a MC time scale which can be obtained by different sets of the simulation
parameters ($\gD N, R_0$). Moreover we show that the scheme is not limited to the heat-bath sampling. 
The limitations of our TQMC development concern mainly the limit of $R_0$  which must be smaller than a threshold 
value obtainable from a trial run. In our calculations, we have got $R_0<0.4$ beyond which we have found deviation 
in the dynamics when compared to those obtained at smaller $R_0$ and the same value of $u_t$.
Notice that, in the framework of the Metropolis sampling, this limits the acceptance rate to $\Ga_{acc}\gtrsim{}0.7$
(see figure~(\ref{relax_ddi_05_6})).

Then we focus
on the frequency dependent susceptibilities, $\chi^{'}(\om)$ and $\chi^{''}(\om)$, as their frequency behavior is well 
known and experimental results for isolated MNP are available. 
We emphasize that our scheme holds also  both for long range dipolar and short range
Heisenberg interactions.
This work gives ways to compare different theoretical
or simulation results on the same models concerning dynamical properties of
MNP. It also helps  to bridge the gap between experimental and simulation results. 

The first application of our TQMC will be to study models of disordered ensembles of MNP where a SG transition is 
obtained from both equilibrium MC simulations and experiments and to compare the transition temperatures obtained 
from the usual equilibrium MC simulations, with the ones obtained from TQMC simulations of AC susceptibilities and 
FC-ZFC dynamical experiments.

\section {Acknowledgements}
 We thank Dr. A. Tamion for providing us with the data relative to the experimental results of
 susceptibility of Ref.~\cite{de-toro_2013b}.
 This work was granted access to the HCP resources of CINES under allocations
 2024-A0160906180 made by GENCI, CINES, France.
 A.M., F.V. and V.R. acknowledge the support of the French Agence Nationale de la Recherche (ANR) through the project
 ANR NanoHype  under grant number ANR-21-CE09-0043.
 J.J.A. thanks SCBI at University of M\'alaga for additional computer time.
  \FloatBarrier
  %
  %
  %
  %
  %
  %
  %
  %
  %
  %
 
  \FloatBarrier
  

\begin{thebibliography}{10}
\bibitem{note_gamma}
We note that in our systems of fixed NP with the moments as unique degree of
  freedom, by using the usual Metropolis sampling in the Monte Carlo simulation
  and imposing a constant acceptance rate through a self consistent adjustment
  of the displacement cone $R$, one obtains as a general rule $R\sim{}\sqrt{T}$
  at least in the non reversible temperature range and suficiently large values
  of the acceptance rate, which corroborate $\Gamma\sim{}cste$ with
  $R=R_0\sqrt{T^*}$.
\end{thebibliography}
  \end   {document}